\documentclass[aps,pra,twocolumn,superscriptaddress]{revtex4-2}
\usepackage[utf8]{inputenc}
\usepackage{soul}
\usepackage{textcomp}
\usepackage[english]{babel}
\usepackage{graphicx}
\usepackage{gensymb} 
\usepackage{dcolumn}
\usepackage{bm}
\usepackage{bm}
\usepackage{amsmath}
\usepackage{verbatim}     
\usepackage[dvipsnames]{xcolor}       
\usepackage[normalem]{ulem}   
\usepackage{bbold}
\usepackage{mathrsfs}
\usepackage[mathscr]{eucal}
\usepackage{physics}      
\newcommand{\qo}[1]{``#1''}                               		
\newcommand{\beq}{\begin{equation}}
\newcommand{\eeq}{\end{equation}}
\newcommand{\bei}{\begin{itemize}}			
\newcommand{\eei}{\end{itemize}}			

\usepackage{hyperref}
\hypersetup{
   pdfpagemode=None,
   pdfstartpage=1,
   pdfmenubar=true,
   pdftoolbar=true,
   colorlinks = true,
   linkcolor=blue,
   citecolor=blue,
   urlcolor=blue,
   bookmarksopen=false
 }
 
\newcommand{\Cite}{\unskip~\cite}
\newcommand{\Eqref}{\unskip~\eqref}
\newcommand{\figref}{\unskip~\ref} 
\newcommand{\fil}[1]{{\color{magenta} #1}}

\usepackage{csquotes}
\begin{document}

\preprint{APS/123-QED}

\title{Universal free-space photonic circuits for polarized structured light}

\author{Dilip Paneru}
\affiliation{%
 Dipartimento di Fisica \qo{Ettore Pancini}, Universit\`{a} degli Studi di Napoli Federico II, 
 Complesso Universitario di Monte Sant'Angelo, Via Cintia, 80126 Napoli, Italy
}%

\author{Paola Savarese}
\affiliation{%
 Dipartimento di Fisica \qo{Ettore Pancini}, Universit\`{a} degli Studi di Napoli Federico II, 
 Complesso Universitario di Monte Sant'Angelo, Via Cintia, 80126 Napoli, Italy
}%

\author{Alessandro Bisio}
 \affiliation{%
Dipartimento di Fisica, Università di Pavia, Via Bassi 6, 27100 Pavia, Italy
 }
\affiliation{Istituto Nazionale di Fisica Nucleare, Sezione di Pavia, Italy}

\author{Francesco Di Colandrea}%
\email{francesco.dicolandrea@unina.it}
\affiliation{%
Dipartimento di Fisica \qo{Ettore Pancini}, Universit\`{a} degli Studi di Napoli Federico II, Complesso Universitario di Monte Sant'Angelo, Via Cintia, 80126 Napoli, Italy
}%

\author{Filippo Cardano}%
\email{filippo.cardano2@unina.it}
\affiliation{%
Dipartimento di Fisica \qo{Ettore Pancini}, Universit\`{a} degli Studi di Napoli Federico II, Complesso Universitario di Monte Sant'Angelo, Via Cintia, 80126 Napoli, Italy
}%

\begin{abstract}
\noindent
Traditional linear-optical schemes for implementing discrete unitary transformations rely on multi-channel interferometers, typically composed of regular meshes of beam splitters and phase shifters. These architectures are well-suited to integrated photonics, where scalability and miniaturization are central requirements. In free space, universal transformations of optical modes have often been pursued through propagation in structured or inverse-designed optical systems, offering advantages in terms of addressable optical modes, reconfigurability, and flexible detection schemes. Here, we present a complementary approach: a collinear free-space photonic circuit that implements the standard interferometric paradigm in a Hilbert space of polarized structured-light modes. The circuit exploits spin-orbit coupling between circular polarization and optical modes carrying quantized transverse momentum, so that conventional beam splitters and phase shifters are replaced by structured optical materials acting directly on selected spin-orbit modes. In our demonstration, liquid-crystal metasurfaces serve as mode beam splitters and phase shifters, and we experimentally realize representative gates in a four-mode space. We further show how the architecture generalizes to larger numbers of modes, providing a scalable route to arbitrary unitary transformations in free space. Numerical optimization over random target unitaries supports the universality of the proposed scheme. These results establish a framework for universal free-space photonic circuits encoded in structured light and structured matter.  
\end{abstract}

\maketitle

\section{Introduction}

Photons are widely regarded as ideal carriers of classical and quantum  information~\cite{krenn2016quantum,Flamini2018,10.1063/1.5115814,mehul_review}, owing to the multiple degrees of freedom available for information encoding, including polarization, time, frequency, and spatial modes, such as orbital angular momentum states. Moreover, their weak interaction with the environment leads to low decoherence, providing an ideal framework for information processing. The use of photons for data processing and transmission lies at the core of optical computing, a paradigm that promises improved performance and lower energy consumption than conventional electronic circuits~\cite{McMahon2023}. In the past few decades, significant technological advances have been made in photon-based computing, with applications ranging from optical neural networks~\cite{Wetzstein2020,ashtiani2022chip,Fu2024} to optimization problems~\cite{Mohseni2022}, quantum computing~\cite{Wang2024a}, and cryptography~\cite{Pai:23}.

A linear-optics implementation of universal operations was first proposed by Reck \emph{et al.}~\cite{reck1994experimental}, demonstrating that any unitary transformation on a discrete set of optical modes can be decomposed into a triangular mesh of tunable two-mode beam splitters and phase shifters. Later, Clements \emph{et al.}~\cite{clements2016optimal} introduced an improved scheme, where the beam splitters are arranged in a rectangular mesh. By design, this scheme achieves minimal optical depth while being more robust to optical losses. In both schemes, reconfigurability is attained by tuning the variable phase elements. In recent years, the demand for customized implementations has fueled intense research activity, leading to the proposal of various schemes analogous to the prototypical architectures, which adapt their underlying principles to alternative physical implementations and design strategies~\cite{saygin2020robust,moralis2023perfect}.

On the implementation side, integrated photonics embeds optical components within planar substrates such as silicon, silicon nitride, or indium phosphide, enabling compact and mechanically stable systems that can be mass-produced using well-established fabrication techniques~\cite{shekhar2024roadmapping}. Furthermore, the new generation of photonic chips allows dynamic programming of the beam splitters' and the optical phase shifters' parameters, in addition to the possibility of programming the mesh geometry itself~\cite{bogaerts2020programmable}. This has led to several demonstrations of on-chip photonic processors capable of manipulating increasingly large numbers of modes~\cite{Taballione:19, Taballione_2021, Taballione2023modeuniversal}. Nevertheless, these platforms face several challenges, most notably fabrication-induced phase errors, waveguide propagation losses, and modal crosstalk arising when circuit elements are densely packed. 

Free-space optical systems provide an alternative approach, offering enhanced flexibility and low propagation losses, especially for large beam sizes and classical bright sources. 
Still, it remains very challenging to scale them beyond the laboratory, as the number of interferometric paths grows quadratically with the number of input modes~\cite{reck1994experimental,PhysRevLett.91.187903}. A frequently adopted solution to mitigate this limitation is the use of diffraction, implemented via an optical Fourier transform or free-space propagation, to make the modes interfere. In this approach, typical of multi-plane light converters (MLPCs), phase masks are usually implemented through spatial light modulators~\cite{Forbes:16}. This scheme can be viewed as the continuous analogue of discrete interferometric meshes~\cite{Martinez-Becerril:24}. Originally introduced for unitary transformations of spatial modes~\cite{morizur2010programmable,Labroille2014}, MPLC schemes have been implemented for a variety of tasks, including mode sorting~\cite{Fontaine2019,PhysRevLett.130.143602} and demultiplexing~\cite{Rouviere:24}. However, the main drawback is the lack of analytical solutions to extract the phase profiles. Therefore, complex optimization techniques are required to configure the system's parameters to maximize fidelity with the target operation~\cite{wang2025}.

The experimental realizations reported thus far have mostly been limited to scalar optical fields. Achieving full control over all photonic degrees of freedom requires extending the technology to also manipulate optical polarization. All spin-orbit technologies~\cite{cardano2015spin,bliokh2015spin}, coupling photons' polarization and spatial structure, are candidates for achieving such functionality. Only recently has a fully vectorial MPLC been implemented using dielectric metasurfaces~\cite{soma2025complete}, reporting a 6-mode vector-beam multiplexer. Nonetheless, a vectorial photonic circuit for optical computing has not yet been demonstrated. 

In this paper, we report an implementation of optical gates simultaneously processing spatial and polarization modes in a four-dimensional (4D) Hilbert space. Spin-orbit interactions are engineered via liquid-crystal metasurfaces (LCMSs), playing the equivalent role of beam splitters and birefringent phase shifters in the original interferometric scheme. Notably, in our framework, the conventional arrays of beam splitters and phase shifters required at each layer are replaced by single liquid-crystal devices. The circuit layout consists of two main repeating units, naturally suited for implementing logical operations in even dimensions. In principle, the number of such units can be scaled up to implement gates in arbitrarily high dimensions. By design, our scheme is $100\%$ efficient, practically limited only by the transmittance of individual devices. Different paradigmatic gates, such as the Hadamard and the CNOT, are demonstrated experimentally. Moreover, we provide strong numerical evidence supporting the claim of universality. Finally, we simulate the effects of experimental imperfections on the setup parameters, observing significant robustness to noise.

\section{Theory}
\subsection{Optical modes and optical circuit}
\label{sec:theory-1}
Our circuit processes circularly polarized optical modes carrying quantized units of transverse momentum. These can be written as~\cite{DErrico:20}
\begin{equation}\label{eqn:opticalmodes}
    \braket{x,y;z}{m,j}= A(x,y,z) e^{i k_z z} e^{i m x\Delta k_{\perp}}\ket{j},
\end{equation}
where $(x,y)$ are the coordinates in the transverse plane, with photons assumed to propagate along $z$. In Eq.\ \eqref{eqn:opticalmodes}, $A(x,y,z)$ is a Gaussian envelope, $k_z$ is the wavevector $z$-component, $\Delta k_{\perp}$ is a unit of transverse momentum, $\ket{j}$ is a circular polarization state, which can be left-handed ${\ket{L}=(1, 0)^T}$ or right-handed ${\ket{R}=(0, 1)^T}$, and $m$ is an integer.

\begin{figure*}[t!]
  \centering
  \hspace*{-0.5cm} 
  \includegraphics[scale=0.54]{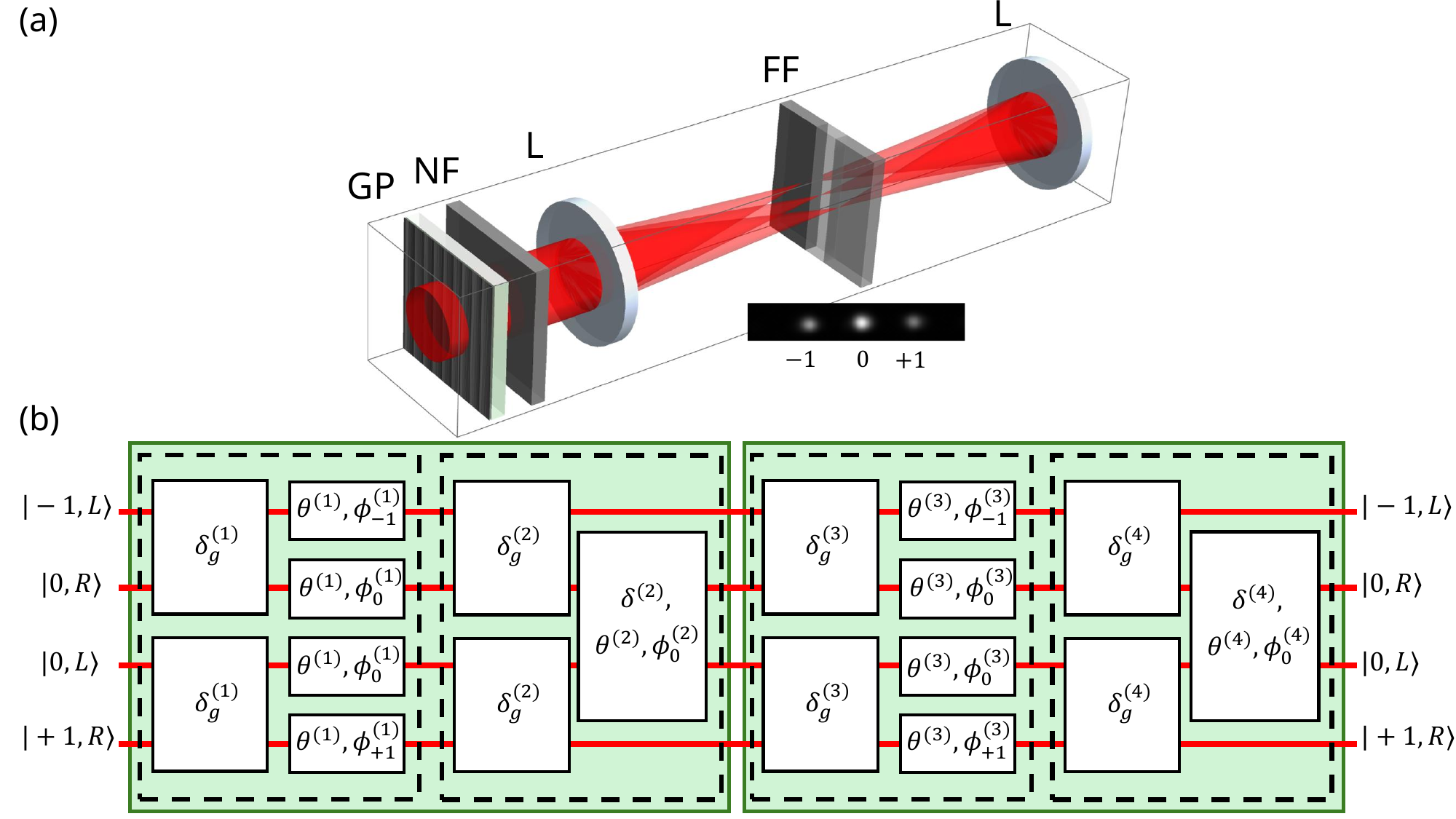}
  \caption{ \textbf{Spin-orbit interferometer for SU(4) optical unitaries.} (a) Inset of a single circuit layer. The modes are processed via a near-field (NF) unit, consisting of a GP and a uniform waveplate, and a far-field (FF) unit with a single LCMS. A lens (L) implements an optical Fourier transform, separating the spatial modes in the focal plane. The FF mask, divided into three patterned sectors, imparts a suitable phase to each mode. 
  (b) Equivalent circuit representation. In the optical implementation, the GPs serve as two-mode beam splitters, while the NF and FF plates act as controllable phase shifters. Green rectangles represent the fundamental two-layer structure that repeats itself to implement arbitrary unitary operations in an even-dimensional Hilbert space. 
  }
  \label{fig:Setup}
\end{figure*}

These modes are manipulated through LCMSs, slabs of nematic liquid crystals artificially patterned on the micrometric scale~\cite{Rubano:19}. More details on the fabrication of LCMSs are provided in Methods. In recent years, these devices have found applications in quantum simulations~\cite{DErrico:20,di2023ultra,ammendola2024large,di2025engineering} and metrology~\cite{Barboza2022}. In particular, LCMSs with periodic modulations of the optic axis can be used to couple the optical modes of Eq.\ \eqref{eqn:opticalmodes} within a range $\pm \ell \Delta k_\perp$, where ${\Lambda/\ell=2\pi/(\ell \Delta k_\perp)}$ is the shortest spatial period in the transverse plane. 
In the circular polarization basis, their Jones matrix reads
\begin{align}
    U(\delta, \theta) = 
    \begin{pmatrix}
    \cos\!\left(\frac{\delta}{2}\right) 
    & i\, e^{-2i\theta(x,y)}\, \sin\!\left(\frac{\delta}{2}\right) \\[6pt]
    i\, e^{2i\theta(x,y)}\, \sin\!\left(\frac{\delta}{2}\right) 
    & \cos\!\left(\frac{\delta}{2}\right)
\end{pmatrix},
\label{eqn:Jones}
\end{align}
where  $\mathcal{\delta} $ is the optical retardation, which can be controlled by applying an AC voltage across the slab~\cite{10.1063/1.3527083}, and $\theta (x,y)$ is the spatially varying optic-axis pattern.
For instance, a $g$-plate~\cite{DErrico:20} is a liquid-crystal polarization grating, where ${\theta(x,y)=\pi x/\Lambda}$. It gives photons momentum shifts of $\pm \Delta k_\perp$ along the $x$ direction, where the shift sign depends on the handedness of the input circular polarization, and can therefore be used as a unitary coupler between neighboring momentum modes with orthogonal polarizations. Remarkably, these modes can be spatially resolved by performing an optical Fourier transform, which is equivalent to measuring the optical field in the focal plane of a lens, where negligible crosstalk is guaranteed as long as ${w_0\geq\Lambda}$, with $w_0$ the Gaussian input beam waist~\cite{DErrico:20}. 

Experimentally, we realize transformations on a 4D Hilbert space, spanned by modes $\lbrace\ket{-1,L},\ket{0,R},\ket{0,L},\ket{1,R}\rbrace$. To implement arbitrary logical operations in SU(4), we cascade four stages of circuit layers. Each layer consists of a near-field (NF) unit, composed of a $g$-plate (GP) and a uniform waveplate, and a far-field (FF) unit, put in the focal plane of a lens, consisting of a single LCMS. The output field is then re-imaged with a second lens as input for the successive layer. The structure of a single circuit layer is shown in Fig.\ \ref{fig:Setup}(a). The main idea behind this scheme is that each GP essentially acts as a tunable beam splitter in the transverse mode space, where the splitting ratio is controlled by the birefringence parameter $\delta_g^{(i)}$ ($i$ labels the layer). The generated modes are separated in the far field, so that each FF unit, patterned as a vertically striped mask, imparts a different phase onto the three spatial modes, given by the mode-dependent optic-axis orientation $\phi_m^{(i)}$. Note that the phase imparted to each mode is equal to twice the optic-axis angle (see Eq.~\eqref{eqn:Jones}). 

Given a target SU(4) unitary, the circuit parameters are found by numerical optimization (see the next section). 
By construction, the birefringence parameters of the first and third NF masks are set to half-wave retardation: 
\begin{equation}
\delta_{NF}^{(1)}=\delta_{NF}^{(3)}=\pi.
\label{eqn:constraint1}
\end{equation}
In the following, we will use ${\delta_{NF}^{(i)}\equiv\delta^{(i)}}$.
At each layer, the FF plates are tuned so as
\begin{equation}
\delta_{FF}^{(i)}=2\pi-\delta^{(i)}. 
\label{eqn:constraint3}
\end{equation}
Moreover, the optic axes of the FF masks in correspondence with the side modes ${m=\pm1}$ are oriented so as
\begin{equation}
\theta^{(i)}=\phi^{(i)}_1=\phi^{(i)}_{-1},
\label{eqn:constraint2}
\end{equation}
for ${i=\lbrace{2,4\rbrace}}$, where $\theta^{(i)}$ is the optic axis of the $i$-th NF uniform plate. The condition enforced by Eq.\ \eqref{eqn:constraint2} allows us to confine the modes within the logical 4D subspace, canceling field contributions associated with the undesired side modes $\ket{-1,R}$ and $\ket{1,L}$, which would couple to higher-order modes through the subsequent GP.
The condition of Eq.\ \eqref{eqn:constraint2} does not need to apply to the first and third mask, since the half-wave retardation (see Eqs.~\eqref{eqn:constraint1}-\eqref{eqn:constraint3}) naturally suppresses the polarization components of the undesired side modes. 

An equivalent circuit representation of our scheme is provided in Fig.~\ref{fig:Setup}(b), where each block acts either as a pairwise mode-coupling operation or a mode-resolved phase manipulation. This picture makes it evident that the fundamental circuit unit that repeats itself is actually made up of two layers (green-shaded rectangles in Fig.~\ref{fig:Setup}(b)), each consisting of a NF and FF unit, suggesting that the circuit is naturally suited for operations in even-dimensional logical spaces. The circuit parameters reported in each operating block are those that undergo optimization to maximize the overlap with a target transformation. 


\subsection{Numerical Optimization}
The circuit parameters are found through numerical optimization. Given a target unitary $U$, we minimize the following cost function: 

\begin{equation}
 \mathcal{L}(\vec{\delta}_{g},\vec{\delta},\vec{\theta},\vec{\phi})=1-F\left(U,U_{\text{opt}}(\vec{\delta}_{g},\vec{\delta},\vec{\theta},\vec{\phi} )\right),
\label{eqn:costfunc}
\end{equation}
where
\begin{equation}
F(U,U_\text{opt})=\dfrac{1}{4}\abs{\text{Tr}\left(U^\dagger U_\text{opt}\right)}
\label{eqn:fidelity}
\end{equation}
is the \emph{gate fidelity} between the target $U$ and the optical circuit $U_\text{opt}$, parametrized in terms of individual LCMSs' settings, for a total of 18 independent parameters. Our optimization protocol consists of successive layers with increasingly complex minimization routines. Starting from a built-in local optimizer, subsequent layers are activated whenever the predefined acceptance criterion is not satisfied (see Methods). For a target unitary, the minimization is accomplished within about 70 ms on average. 

\begin{figure}[t!]
  \centering
   \hspace*{0cm} 
  \includegraphics[scale=0.3]{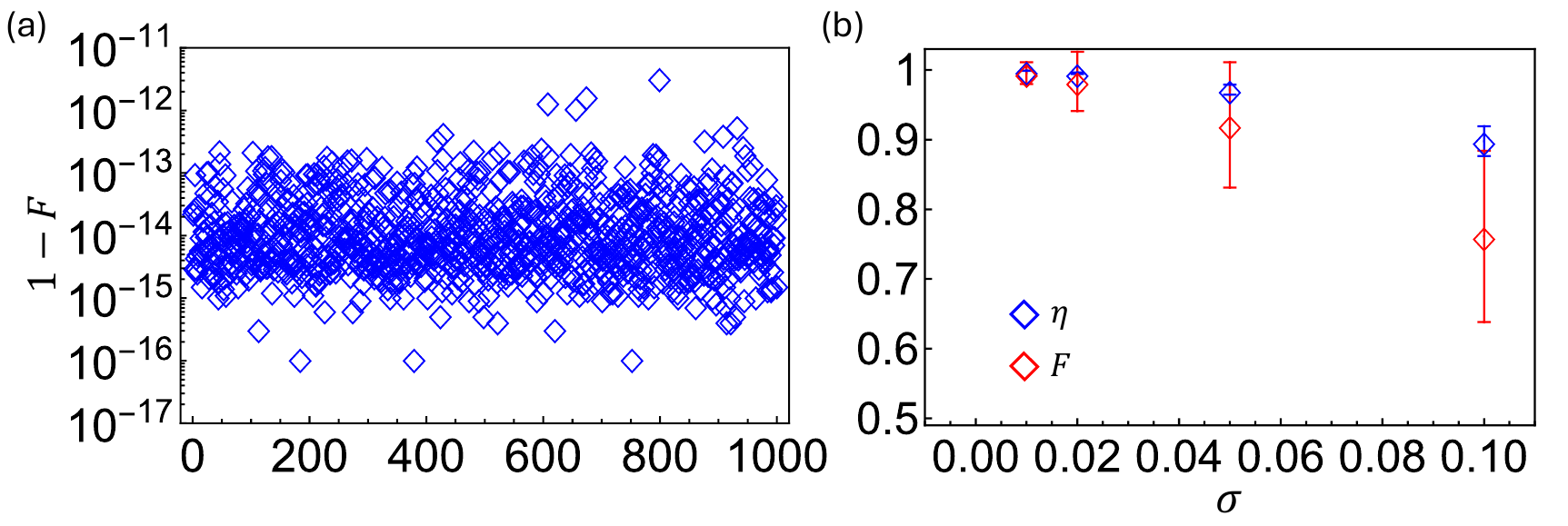}
  \caption{\textbf{Numerical optimization.} (a) Log-plot of the infidelity, $1-F$, obtained for 1000 Haar-random SU(4) unitaries after the optimization of the circuit parameters. (b)~Average efficiency and fidelity for varying levels of simulated Gaussian noise on all the setup parameters.  Error bars are computed as the standard deviation over the set of 1000 unitaries.
  }
  \label{fig:NoiseSim}
\end{figure}

Universality in the 4D space is supported by numerical optimization. In particular, we run the optimization over a set of 1000 Haar-random SU(4) unitaries, obtaining an average infidelity (see Eq.\ \eqref{eqn:costfunc}) of ${3.5\cdot10^{-14}}$. This certifies excellent optimization performance and demonstrates the ability of the circuit to implement arbitrary unitaries. The infidelity obtained for each target unitary is plotted in Fig.\ \ref{fig:NoiseSim}(a). 

We argue that the circuit architecture can be scaled to implement operations in SU($2n$) for arbitrarily large integers $n$, by cascading $n$ stages of the two-layer core scheme (see Fig. \figref{fig:Setup}(b)). This large-scale inference is supported by the close analogy between our spin-orbit setup and the traditional interferometric schemes for implementing discrete unitary transformations~\cite{reck1994experimental,clements2016optimal}. The two schemes indeed share the same conceptual framework, with the main difference being that, in our platform, interference occurs in mode space among copropagating optical modes rather than between spatially separated optical paths. Of course, a realistic $n$-layer experimental implementation will face relevant challenges, most notably alignment and optical losses. 

Finally, we test the robustness of our circuit to different levels of experimental noise. To do so, for each of the 1000 unitaries, we assume that all the setup parameters, both the ones that are fixed by design (see Eqs.~\eqref{eqn:constraint1}-\eqref{eqn:constraint2}) and the ones obtained from the optimization, are affected by a relative error extracted from a zero-mean Gaussian distribution, with standard deviation equal to $1\%$, $2\%$, $5\%$, and $10\%$ of the nominal values. For every noise level, we run 50 numerical experiments and, for each unitary, we compute the obtained efficiency (the fraction of output light coupled to the logical subspace) and gate fidelity (after post-selecting on the logical modes and renormalizing the output light intensity coupled to them), extracted by averaging over all simulated realizations. The results obtained by averaging over all the unitaries are illustrated in Fig.\ \ref{fig:NoiseSim}(b). These show that the circuit is very robust to moderate perturbations up to $5\%$ relative errors, but can also tolerate stronger errors of the order of $10\%$, for which the average efficiency and fidelity drop to $0.90\pm0.02$ and $0.76\pm0.12$, respectively.

\begin{figure*}[t!]
  \centering
   \hspace*{-0.45cm} 
  \includegraphics[scale=0.32]{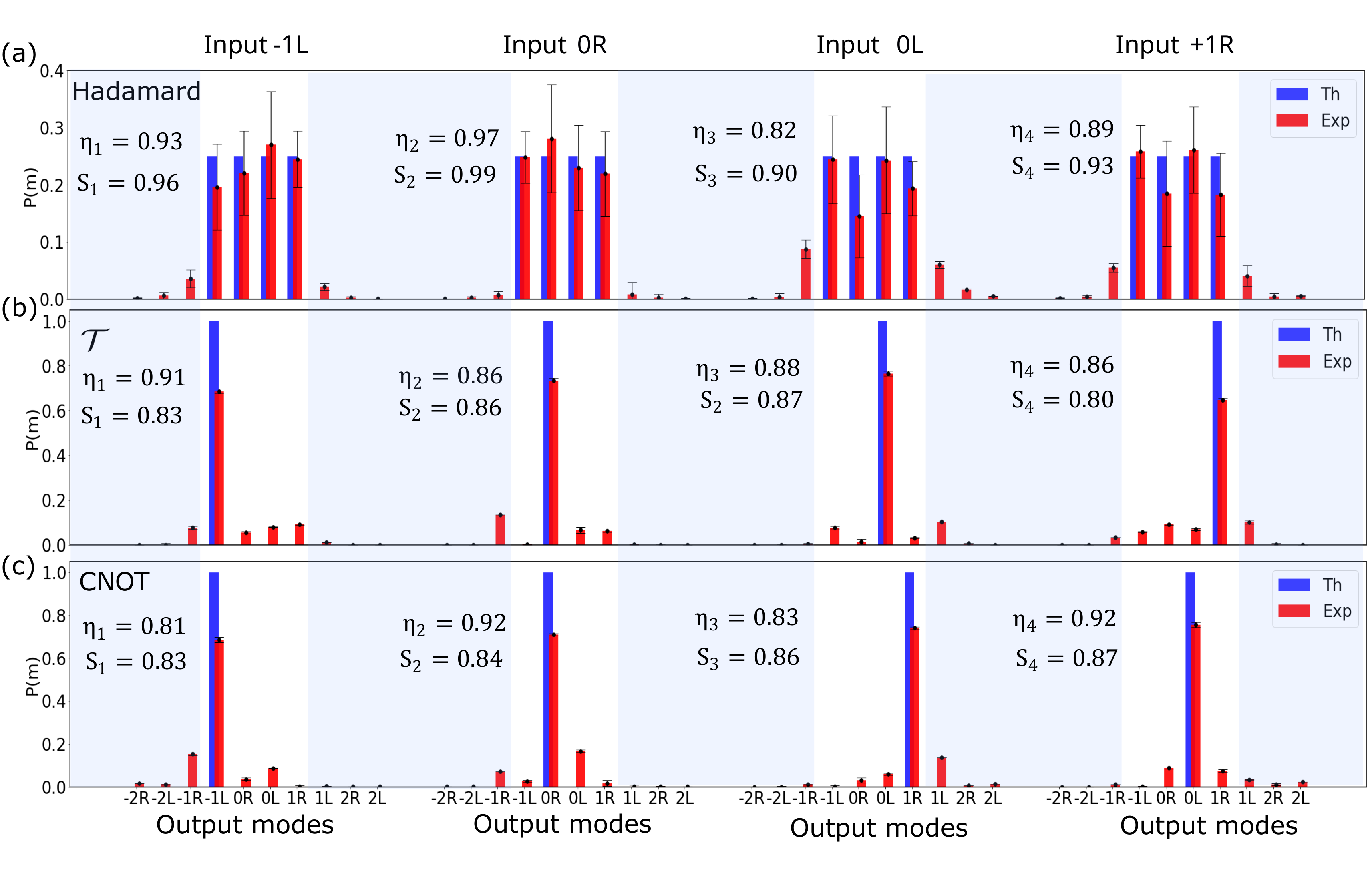}
   \vspace*{-0.75cm} 
  \caption{\textbf{Experimental probability distributions.} Experimental (red) probability distributions for the (a) Hadamard, (b) $\mathcal{T}$, and (c) CNOT gate (see main text for the definition), obtained for each input state and compared with theoretical predictions (blue). Experimentally, a fraction of the output light couples to modes outside the target 4D Hilbert space. The blue-shaded regions of each plot correspond to the extraneous modes. Average efficiency and average similarity over the four input states are ${\bar{\eta}=0.90, 0.88, 0.87}$ and ${\bar{S}=0.95\pm0.08,0.84\pm0.01,0.85\pm0.01}$, respectively. Error bars are obtained from Monte Carlo simulations, as detailed in the main text. 
  }
  \label{fig:ExpResultsSI}
\end{figure*}

\section{Results}
We experimentally validate our circuit using a classical laser source. A He-Ne laser beam (operating wavelength: ${\lambda \simeq 633}$~nm) is expanded so that the input beam waist is approximately ${w_0 \simeq 1.3}$ mm.  As discussed in Sec.~\ref{sec:theory-1}, this is sufficient to clearly separate the modes in the far field, as the fabricated GPs have a spatial period ${\Lambda=1}$~mm. Any input state from the logical basis can be prepared via a combination of a quarter-wave plate (QWP) and a GP, by adjusting the QWP angle and the GP birefringence parameter. The action of the circuit on the beam corresponds to a target SU(4) transformation of the logical modes. Within each layer, a lens with focal length ${f=15}$~cm connects the near field to the far field via an optical Fourier transform. In the focal plane, a LCMS is patterned with ${93\text{-}\mu\text{m}}$-wide vertical stripes, corresponding to the modes' spatial separation. A second 15-cm lens is then used to image the output field onto the subsequent layer. After the four layers, polarization projections are performed using a QWP and a linear polarizer to separately resolve the left and right components of the output state, and the resulting spatial-mode distribution is recorded with a CCD camera, put in the image plane of the last FF mask. Each output mode appears as a Gaussian spot, covering approximately 20 pixels. The experimental probability distribution ${P^\text{exp}(m)}$ is thus extracted by integrating the light intensity within each spot and normalizing it to the total intensity.  

For our demonstration, we implement three paradigmatic gates, specifically the 4D Hadamard,
\begin{equation}
H=\frac{1}{2}\begin{pmatrix}
1 & 1 & 1 & 1\\
1 & -1 & 1 & -1\\
1 & 1 & -1 & -1\\
1 & -1 & -1 & 1
\end{pmatrix},
\end{equation}
the operator ${\mathcal{T}=I\otimes T}$, where $I$ is the ${2\times 2}$ identity and 
\begin{equation}
T=\begin{pmatrix}
1 & 0\\
0 & e^{i\pi/4}
\end{pmatrix}
\end{equation}
is the \qo{magic} gate \Cite{PhysRevLett.128.050402}, and the CNOT gate,
\begin{equation}
\text{CNOT}=\begin{pmatrix}
1 & 0 & 0 & 0\\
0 & 1 & 0 & 0\\
0 & 0 & 0 & 1\\
0 & 0 & 1 & 0
\end{pmatrix}.
\end{equation}
Interestingly, these operators, along with their equivalent parametrizations or representations, constitute a universal gate set for fault-tolerant quantum computation~\cite{shor,doi:10.1142/9789812385253_0008}.

\begin{figure*}[th!]
  \centering
   \hspace*{-0.55cm} 
  \includegraphics[scale=0.32]{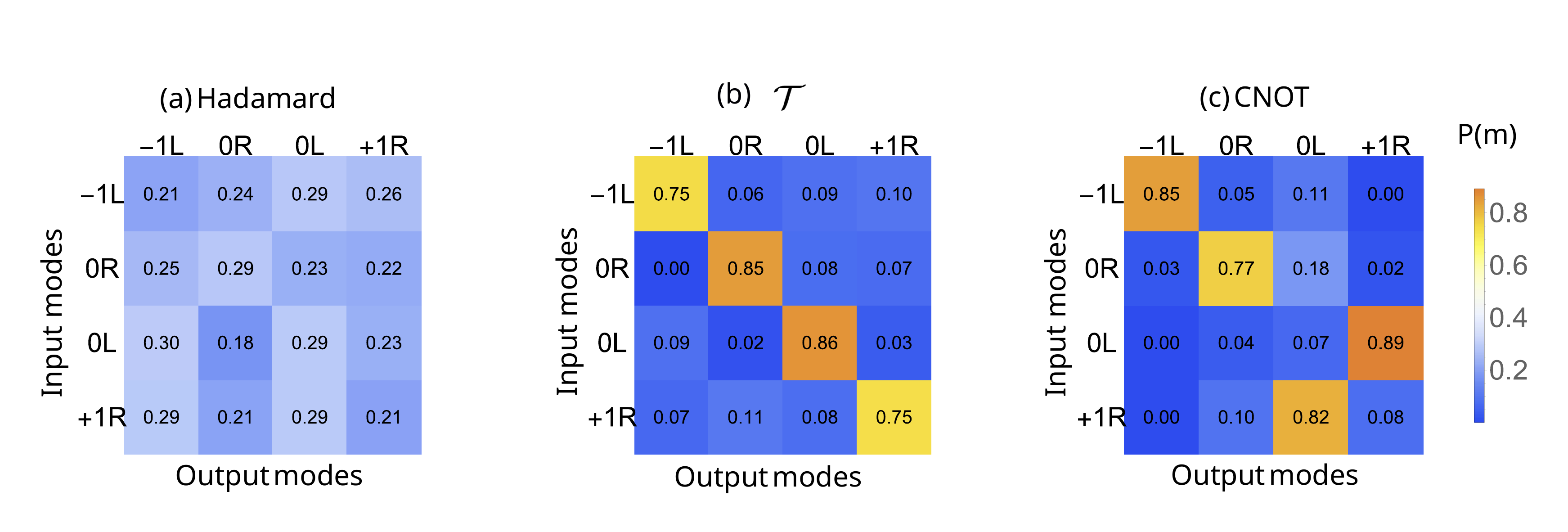}
  \vspace*{-0.75cm} 
  \caption{\textbf{Experimental gate probabilities after subspace renormalization.} Experimental renormalized probability distributions obtained for the (a) Hadamard, (b) $\mathcal{T}$, and (c) CNOT gates. The matrix entries are the experimental probability distributions obtained by preparing the four logical input states, corresponding to the matrix rows, and renormalizing the amount of light coupled only to the desired 4D output Hilbert space.
  }
  \label{fig:ExpResults}
\end{figure*}

In Fig.\ \ref{fig:ExpResultsSI}, for each input state, we report the probability distributions obtained experimentally for the three implemented operations, comparing them with theoretical predictions. By construction (see Eqs.\ \eqref{eqn:constraint1}-\eqref{eqn:constraint2}), our circuit has unit efficiency. However, experimental imperfections drive it away from the ideal realization. For this reason, we also record a fraction of the output light coupled to the extraneous modes $ \lbrace{\ket{-2,R},\ket{-2,L},\ket{-1,R},\ket{1,L},\ket{2,R}\ket{2,L}\rbrace}$, as shown in Fig.\ \ref{fig:ExpResultsSI}. We then quantify the experimental efficiency $\eta$ of the circuit, defined as the fraction of the output light coupled to the target 4D space. The efficiency associated with each input state is reported in Fig.\ \ref{fig:ExpResultsSI}, from which we also extract the average efficiency $\bar{\eta}$. For the implemented gates, we obtain ${\bar{\eta}=0.90, 0.88,}$ and ${0.87}$, respectively. From the probability distributions over the extended Hilbert space, we compute the \emph{similarity},
\begin{equation}
S_i=\sum_{m} \sqrt{P^\text{th}_{i}(m)P^\text{exp}_{i}(m)}\,,
\end{equation}
for the $i$-th input state, where the output-mode index $m$ runs over the extended Hilbert space. The similarity associated with each input state is reported in Fig. \figref{fig:ExpResultsSI} for each gate. The average values over the four input states are ${\bar{S}=0.95\pm0.08, 0.84\pm0.01,}$ and ${0.85\pm0.01}$, respectively, denoting a good agreement with the theory. These results confirm that the circuit implements the target operation with high accuracy. Deviations from the theory are mainly ascribed to fabrication defects and thermal fluctuations, affecting the liquid-crystal electric response. Besides, the similarity metric tends to penalize sparse output distributions expected to feature single peaks, partially explaining the lower values obtained for the $\mathcal{T}$ and CNOT gates. Error bars in Fig. \figref{fig:ExpResultsSI} are obtained from Monte Carlo simulations, where the optic-axis angles and the birefringence of each NF and FF LCMSs are assumed to be Gaussian random variables with $1\%$ and $5\%$ standard deviations, respectively, estimated through direct measurements on the fabricated devices.

Since residual coupling to undesired modes is observed experimentally, we further process the experimental data only within the target output-mode subspace. The measured probabilities are therefore post-selected on this subspace and renormalized accordingly. The resulting probabilities should thus be interpreted as conditional on detection in one of the target modes spanning the desired 4D space. Applying this renormalization procedure yields the output probabilities reported in Fig. \figref{fig:ExpResults} for each logical input state, corresponding to the matrix rows. The obtained matrices show that, despite experimental losses, the desired logical functionality of each implemented gate is reproduced with high accuracy within the target subspace. Here, we do not explicitly compute the gate fidelity (see Eq. \Eqref{eqn:fidelity}), as this would also require extracting the phase information of the measured modes. However, this information could be obtained from standard interferometry or measurements in conjugate planes~\cite{DiColandrea2024}.

\section{Discussion}
We have demonstrated a free-space optical circuit for the implementation of SU(4) unitaries. The first remarkable achievement is that our circuit satisfies the key requirement of (ideally) ${100\%}$ efficiency. This feature becomes especially critical in the single-photon regime and for scaling this approach to higher dimensions. 

Universality is the second core claim of the present work. Although a complete analytical proof is currently lacking, numerical optimizations performed on Haar-random unitaries provide strong evidence that the circuit is capable of implementing arbitrary transformations with (ideally) $100\%$ fidelity. This argument is further supported by a direct mapping between the implemented circuit elements and the standard building blocks employed in universal linear-optical decompositions of discrete unitary operators~\cite{reck1994experimental,clements2016optimal}. This mapping offers a clear physical interpretation of our architecture and justifies the claim. Additionally, in our implementation, the typical array of beam splitters and phase elements at each layer is encoded within a single device for all the modes. Since the number of layers is expected to increase linearly with the number of processed modes, this also implies only a linear increase in the fabrication complexity, compared to traditional approaches.

Our results establish a novel paradigm for optical computing based on spin-orbit photonic gadgets working in free space, enabling fully vectorial control of the light structure. In our demonstration, we used liquid-crystal metasurfaces, but the same functionality could also be achieved with other spin-orbit devices, such as dielectric metasurfaces or other metamaterials~\cite{capasso1,capasso2,capasso3,soma2025complete}.\\A natural generalization is to extend our approach to modes arranged on a two-dimensional grid rather than on a line. This will require patterning the devices in both the $x$ and $y$ directions to fully exploit the two-dimensional nature of the transverse momentum of light~\cite{DErrico:20,ammendola2024large}. The use of the second spatial dimension can also enable multiplexing different unitary operations, by patterning the optic axes of the near- and far-field masks as successive vertically displaced liquid-crystal stripes.\\Another interesting avenue is engineering optical losses to expand the computational capabilities of the circuit toward the realization of higher-dimensional unitaries via post-selection methods or non-unitary gates~\cite{Verstraete2009,Harrington2022}. \\Additionally, one can characterize the statistical distributions of the circuit parameters so that the induced measure on the implemented transformations matches the Haar measure~\cite{Burgwal:17}. This would enable direct sampling of Haar-random unitaries from the circuit parametrization, as required for applications such as boson sampling~\cite{PhysRevLett.119.170501,adv_photonics_review}.

Finally, we plan to extend the circuit concept and its validation to two-photon states. For this purpose, the availability of new detection systems, such as SPAD arrays~\cite{Makowski:24} or intensified single-photon cameras~\cite{Nomerotski_2023}, will significantly boost the reconstruction of spatially structured quantum states of light. 

\section*{Methods}
\subsection{Fabrication}
Liquid-crystal metasurfaces are birefringent slabs of nematic liquid crystals, whose molecular director is patterned on the micrometric scale. The fabrication process starts by cleaning two glass plates, precoated with a thin film of indium tin oxide (ITO), a transparent conductive material, in an ultrasonication chamber, first with a 5\% solution of phosphate-free alkaline detergent and distilled water, then with distilled water only.  Both cleaning steps are performed at ${60^\circ \text{C}}$ for 1 hour. The plates are then dried in an oven at $100^\circ\text{C}$ for 1 hour. This is followed by a 45-min Ozone-UV exposure to promote surface activation and facilitate azo-dye deposition. The glass plates are spin-coated with a thin film of photosensitive azo-dye. After spin-coating, the plates are sandwiched together using $6$-$\mu$m silica spacers, defining the cell thickness. The desired pattern is imprinted using a well-established photoalignment technique~\cite{Rubano:19}, driving the azo-dye molecules to orient point by point following the polarization of an incident linearly polarized beam at $445$~nm. After photoalignment, the cell is filled with nematic liquid crystals, which align with the azo-dye substrate and penetrate by capillarity the sample pre-heated at $100^\circ\text{C}$.

\subsection{Optimization Routine}
The minimization pipeline follows a hierarchical optimization scheme composed of successive stages, implemented in Wolfram Mathematica. For a target unitary, a solution is considered acceptable if the cost function defined in Eq.~\eqref{eqn:costfunc} is minimized to a value below $10^{-12}$. An initial minimization attempt is performed using the routine FindMinimum, a built-in local optimizer. When the obtained solution does not satisfy the acceptance criterion, a second run of the same optimizer is executed by increasing the maximum number of allowed iterations. Finally, if convergence is still not achieved, the global optimizer NMinimize is employed to find the global minimum, and its output is retained as the final solution. 
\\
\section*{Acknowledgements}
DP, AB, and FC acknowledge financial support from European Union–Next Generation EU through the MUR project Progetti di Ricerca d'Interesse Nazionale (PRIN) DISTRUCT No.~P2022T2JZ9. FDC acknowledges financial support from the European Union–Next Generation EU through the PNRR MUR Project No.~PE0000023-NQSTI.
\bibliography{bibliography}

\end{document}